\documentclass[epj]{svjour}
\usepackage{graphics}
\begin{document}
\title{A fully relativistic description of Hypernuclear production in proton- 
and pion-Nucleus Collisions}
\author{R. Shyam\inst{1}
\thanks{\emph{Present address:Institut f\"ur Theoretische Physik, Universit\"at
Giessen, D-35392 Giessen}} 
}                    
\institute{Saha Institute of Nuclear Physics, 1/AF Bidhan Nagar, Kolkata, India}
\date{Received: date / Revised version: date}
%
\abstract{
Exclusive $A(p,K^+){_\Lambda}B$ and $A(\pi,K^+){_\Lambda}B^\prime$ reactions
leading to two body final states, have been investigated in a fully
covariant model based on an effective Lagrangian picture. The explicit
kaon production vertex is described via creation, propagation and decay
into relevant channel of $N^*$(1650), $N^*$(1710) and $N^*$(1720) 
intermediate baryonic resonance states, in the initial collision of the 
projectile with one of the target nucleons.
The bound state wave functions are obtained by 
solving the Dirac equation with appropriate scalar and vector potentials.
The calculated cross sections show strong sensitivity to the final 
hypernuclear state excited in the reaction. Cross sections of 1 - 2 nb/sr
are obtained at peak positions of favored transitions in case of
the $A(p,K^+){_\Lambda}B$ reaction on heavier targets.
} 
 
\maketitle
\section{Introduction}
\label{intro}
$\Lambda$ hypernuclei have been studied extensively by stopped
as well as in-flight $(K^-,\pi^-)$~\cite{chr89,ban90} and 
$(\pi^+,K^+)$ reactions \cite{has06}. The kinematical properties
of the $(K^-,\pi^-)$ reaction allow only a small momentum transfer to
the nucleus (at forward angles), thus there is a large probability of
populating $\Lambda$-substitutional states in the residual hypernucleus.
On the other hand, in the $(\pi^+,K^+)$ reaction the momentum transfer
is larger than the nuclear Fermi momentum, therefore, hypernuclear states
with configurations of an outer neutron hole and a $\Lambda$ hyperon in a
series of orbits covering all bound states can be excited in this case.
The data on the hypernuclear spectroscopy have been used 
extensively to extract information about the hyperon-nucleon interaction 
within a variety of theoretical approaches (see, e.g.,~\cite{hiy00,kei00}).

Alternatively, $\Lambda$-hypernuclei can also be produced with high 
intensity proton beams via the $(p,K^+){_{\Lambda}}B$ reaction where the
hypernucleus ${_{\Lambda}}B$ has the same neutrons and proton numbers as
the target nucleus $A$.  First set of data for this reaction on deuterium
and helium targets, have already been reported in Ref.~\cite{kin98}. 
The states of the hypernucleus excited in the $(p,K^+)$ reaction may have
a different type of configuration as compared to those excited in the
$(\pi^+,K^+)$ reaction. Thus a comparison of informations extracted from
the study of two reactions is likely to lead to a better
understanding of the hypernuclear structure.

Theoretical studies of the $A(p,K^+){_{\Lambda}}B$ reaction
are based on two main approaches; the one-nucleon model (ONM)
[Fig.~1(a)] and the two-nucleon model (TNM) [Fig.~1(b) and 1(c)].
In the ONM the incident proton first scatters from the target nucleus
and emits a (off-shell) kaon and a $\Lambda$ hyperon which gets captured
into one of the (target) nuclear orbits. Thus there is only one 
single active nucleon (impulse approximation) which carries the entire 
momentum transfer to the target nucleus (about 1.0 GeV/c at the outgoing
$K^+$ angle of 0$^\circ$ ). This makes this model  extremely sensitive 
to details of the bound state wave function at very large momenta where
its magnitude is very small leading to quite low cross sections. In the
ONM calculations of $(p,K^+)$ and also of $(p,\pi)$ reactions the 
distortion effects in the incident and the outgoing channels have been
found to be quite important~\cite{kri95,coo82}.

In the two-nucleon mechanism (TNM), the
kaon production proceeds via a collision of the projectile nucleon
with one of its target counterparts. This excites intermediate baryonic
resonant states which decay into a kaon and a $\Lambda$ hyperon.
The nucleon and the hyperon are captured into the respective nuclear
orbits while the kaon rescatters into its mass shell. In this picture 
there are altogether three active bound state baryon wave functions
taking part in the reaction process allowing the large momentum
transfer to be shared among three baryons. Consequently, the
sensitivity of the model is shifted from high momentum to the
lower momentum parts of the bound state wave functions where they
are relatively better known.
\begin{figure}
\begin{center}
\resizebox{0.35\textwidth}{!}{
\includegraphics{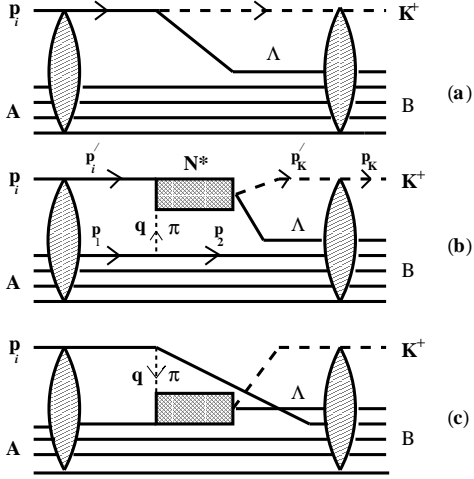}
}
\end{center}
\caption{Graphical representation of one- and two-nucleon models of
$A(p,K^+){_{\Lambda}}B$ reaction.
The elliptic shaded area represent the optical model interactions in
the incoming and outgoing channels.
}
\end{figure}

A fully covariant TNM for the $A(p,K^+){_{\Lambda}}B$ reaction has 
been developed in Refs~\cite{shy04,shy06} 
by retaining the field theoretical structure of the
interaction vertices and by treating the baryons as Dirac particles.
The initial interaction between the incoming proton and a bound nucleon of
the target is described by $\pi$, $\rho$ and $\omega$ exchange mechanisms.
This leads to the excitations of $N^*$(1650)[$\frac{1}{2}^-$],
$N^*$(1710)[$\frac{1}{2}^+$], and $N^*$(1720)[$\frac{3}{2}^+$] baryonic
resonance intermediate states which decay into kaon and the $\Lambda$
hyperon. We present a brief sketch of this formalism in the next section. 
\section{Covariant Two-nucleon Model}

We provide here only a brief outline of our TNM; details of this theory
are given in Refs.~\cite{shy04,shy06}. The structure of the TNM for the 
$(p,K^+)$ reaction is similar to that of the effective Lagrangian 
approach for the elementary $pp \to p\Lambda K^+$ process as discussed in
Ref.\cite{shy99}. We use the same effective Lagrangians and vertex 
parameters to model the initial interaction between the incoming proton 
and a bound nucleon of the target by means of $\pi$, $\rho$ and $\omega$
exchange mechanisms. The structure for the resonance-nucleon-meson vertices
were also taken to be the same. Terms corresponding to the interference 
between various amplitudes are retained. After having established the 
effective Lagrangians and the coupling constants for  various vertices the
amplitudes for the graphs 1b and 1c can be written by following the well 
known Feynman rules. These amplitudes can be evaluated by following the 
techniques described in Ref.~\cite{shy06}. It should be noted that the
Fig. 1c can be used in a straight forward way to calculate the amplitudes of
the $A(\pi,K^+){_\Lambda}B^\prime$ reaction.

The differential cross section for the $(p,K^+)$ reaction is given by
\begin{eqnarray}
\frac{d\sigma}{d\Omega} & = & \frac{1}{(4\pi)^2}
\frac{m_pm_Am_B}{(E_{p_i} + E_A)^2} \frac{p_K}{p_i}
 \sum_{m_im_f} |T_{m_im_f}|^2,
\end{eqnarray}
where $E_{p_i}$ and $E_A$ are the total energies of incident
proton and the target nucleus, respectively while $m_p$, $m_A$ and $m_B$ are
the masses of the proton, and the target and residual nuclei, respectively.
The summation is carried out over initial ($m_i$) and final ($m_f$)
spin states.  $T$ is the final $T$ matrix obtained by summing the transition

\label{sec:1}
\section{Results and Discussions}
\label{sec:2}
The spinors for the final bound hypernuclear state (corresponding to
momentum $p_\Lambda$) and for two intermediate nucleonic states
(corresponding to momenta $p_1$ and $p_2$) are required to
perform numerical calculations of various amplitudes. We assume these
states to be of pure-single particle or single-hole configurations
with the core remaining inert. The spinors in the momentum space are
obtained by Fourier transformation of the corresponding coordinate space
spinors which are the solutions of the Dirac equation with potential fields
consisting of an attractive scalar part ($V_s$) and a repulsive vector part
($V_v$) having a Woods-Saxon form. The potential parameters are given in 
Ref.~\cite{shy06}.
\begin{figure}
\begin{center}
\resizebox{0.30\textwidth}{!}{
\includegraphics{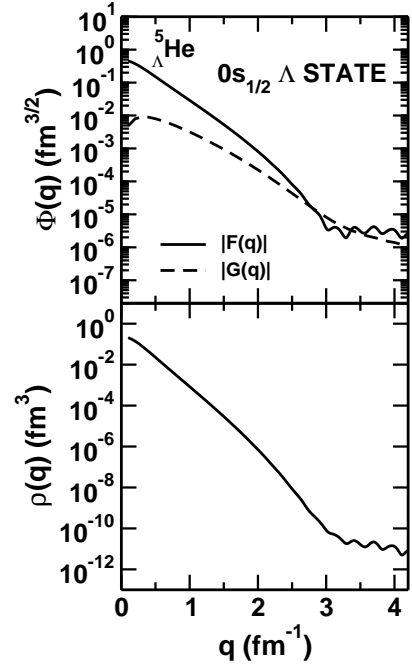}
}
\end{center}
\caption{(Upper panel) Momentum space spinors [$\Phi(q)$] for
$0s_{1/2}$ $\Lambda$ orbit in $^{5}\!\!\!_\Lambda$He hypernucleus.
$|F(q)|$ and $|G(q)|$ are the upper and lower components
of the spinor, respectively. (Lower panel) Momentum distribution
[$\rho(q) = |F(q)|^2 + |G(q)|^2$] for
the same hyperon state calculated with these wave functions.
}
\end{figure}

In Figs.~2 and 3, we show the lower and upper components of the Dirac 
spinors in momentum space for the $0s_{1/2}$ hyperon in
$^{5}\!\!\!_\Lambda$He  and the $0p_{3/2}$ and $0s_{1/2}$ hyperons in
$^{13}\!\!\!_\Lambda$C, respectively. In each case, we note that only
for momenta $<$ 1.5 fm$^{-1}$, is the lower component of the spinor 
substantially smaller than the upper component. In the region of 
momentum transfer pertinent to exclusive kaon production 
the lower components of the spinors are not negligible as 
compared to the upper component which demonstrates that a fully
relativistic approach is\\ essential for an accurate description of this 
reaction. The spinors calculated in this way provide a good description 
of the experimental nucleon momentum distributions for various nucleon 
orbits as is shown in Ref.~\cite{shy95}. In the lower panel of each of
Figs.~2 and 3 we show momentum distribution $\rho(q)\,[ = F(q)|^2 + |G(q)|^2]$
of the corresponding $\Lambda$ hyperon. In each case the momentum density of
the hyperon shell, in the momentum region around 0.35 GeV/c, is at least 2-3
orders of magnitude larger than that around 1.0 GeV/c. In the TNM, therefore
cross section will be larger than the those in the ONM as in the former
case the sensitivity of the model gets shifted to lower momenta due to
involvement of three baryon in the momentum sharing process.
\begin{figure}
\begin{center}
\resizebox{0.40\textwidth}{!}{
\includegraphics{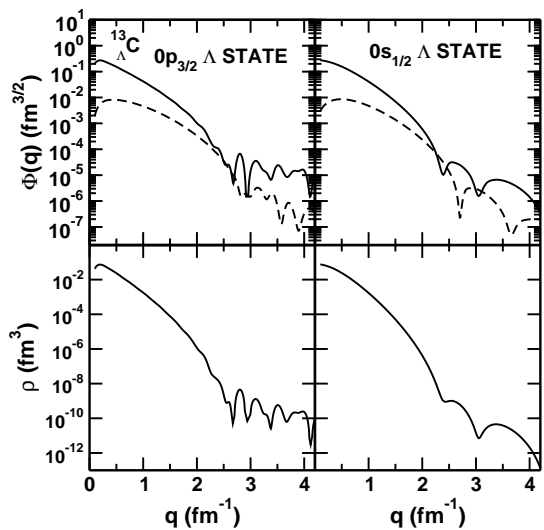}
}
\end{center}
\vskip 0.10in
\caption{same as in Fig.~2 for $0p_{3/2}$ $\Lambda$ and
$0s_{1/2}$ $\Lambda$ orbits in hypernucleus
$^{13}\!\!\!_\Lambda C$. 
}
\end{figure}

A crucial quantity needed in the calculations of the kaon
production amplitude is the pion self-energy, $\Pi(q)$,
which takes into account the medium effects on the intermediate pion
propagation. Since the energy and momentum carried by such a pion
can be quite large (particularly at higher proton incident energies),
a calculation of $\Pi(q)$ within a relativistic approach is mandatory.
In our study contributions from the particle-hole 
($ph$) and delta-hole ($\Delta h$) excitations are taken into account.
The self-energy has been renormalized by including the short-range repulsion
effects by introducing the constant Landau-Migdal parameter
$g^\prime$ which is taken to be the same for $ph-ph$ and $\Delta h-ph$
and $\Delta h-\Delta h$ correlations which is a common choice.
\begin{figure}
\begin{center}
\resizebox{0.35\textwidth}{!}{
\includegraphics{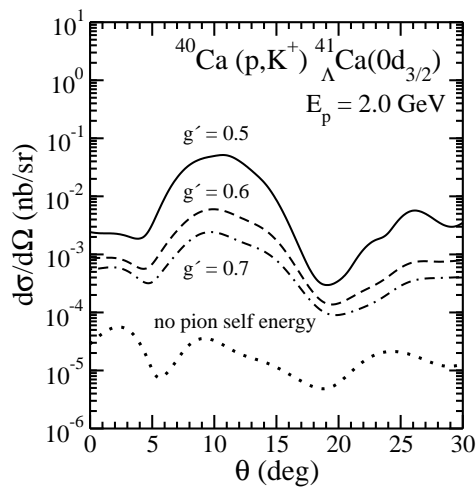}
}
\end{center}
\caption{Differential cross section for the
$^{40}$Ca$(p,K^+)$$^{41}\!\!\!_\Lambda$Ca($0d_{3/2}$)
reaction for the incident proton energy of 2.0 GeV.
The dotted line shows the results obtained without including the pion
self-energy in the denominator of the pion propagator while
full, dashed and dashed-dotted lines represent the same calculated
with pion self-energy renormalized with Landau-Migdal parameter of
0.5, 0.6 and 0.7, respectively.
}
\end{figure}

In Fig.~4, we show the dependence of our calculated cross sections on
pion self-energy.  It is interesting to note that this has a rather
large effect. We also note a surprisingly large effect on the
short range correlation (expressed schematically by the Landau- Migdal
parameter $g^\prime$). Similar results have also been reported in
case of the $(p,\pi)$ reactions. However, more definite
statements about the usefulness of $(p,K^+)$ reactions in
probing the medium effects on the pion propagation in nuclei
must await the inclusions of distortions in the incident and outgoing
channels. 
 
In Figs.~5 and 6, we show the kaon angular distributions 
for various final hypernuclear states excited in 
$^{12}$C$(p,K^+)$$^{13}\!\!\!_\Lambda$C and 
$^{4}$He$(p,K^+)$$^{5}\!\!\!_\Lambda$He reactions, respectively. The incident
proton energies in the two cases are taken to be 1.8 GeV and 2.0 GeV,
respectively where the angle integrated cross sections for the two
reactions are maximum. The calculations are the coherent sum of all 
the amplitudes corresponding to the various meson exchange processes 
and intermediate resonant states. Clearly, the cross sections are quite
selective about the excited hypernuclear state, being maximum for the
state of largest orbital angular momentum. This is due to the large 
momentum transfer involved in this reaction.

It should be noted that in 
in Fig.~2 the angular distribution has a maximum at angles larger
than 0$^\circ$. In contrast to this, the maximum in Fig~3, occurs at 
zero degree and it decreases gradually as the angle increases.
\begin{figure}
\resizebox{0.40\textwidth}{!}{
\includegraphics{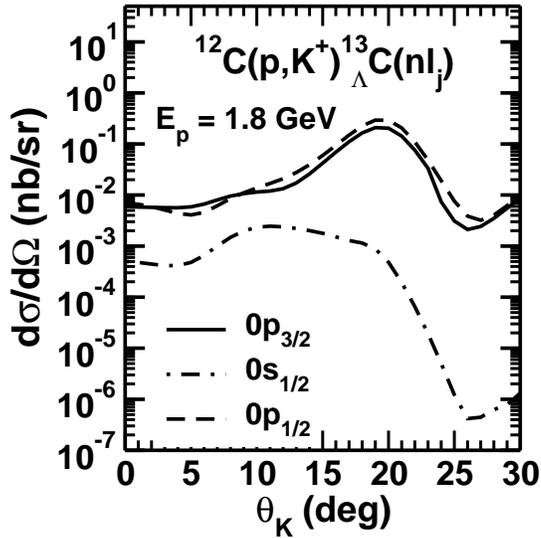}
}
\caption{Differential cross section for the
$^{12}C(p,K^+)^{13}\!\!\!_\Lambda C$
reaction for the incident proton energy of 1.8 GeV for various
bound states of final hypernucleus as indicated in the figure.
}
\end{figure}
This is  due to the fact that in the bound state spinors of 
$^{13}\!\!\!_\Lambda$C, there are several maxima in the upper
and lower components of the momentum space bound spinors in the
region of large momentum transfers. Therefore, in the kaon angular
distribution the first maximum may shift to larger angles
reflecting the fact that the bound state wave functions show
diffractive structure at higher momentum transfers. On the other
hand, for momentum transfers relevant to 
$^{4}$He$(p,K^+)$$^{5}\!\!\!_\Lambda$He reaction the Dirac spinors are
smoothly varying and are devoid of structures as can be seen in Fig.~2.
In any case, from the purely kinematical arguments it is clear that the
maximum in the cross section for a $\ell = 0$ transition is expected
to occur at smaller angles as compared that for a $\ell \neq 0$ one.

\section{Summary and Conclusions}

\label{sec:3}

In summary, we have made a study of the $A(p,K^+){_\Lambda}B$ reaction
on  $^{4}$He, $^{12}$C, and $^{40}$Ca targets and of the$A(p,K^+){_\Lambda}B$ 
reaction on the $^{12}$C target within a fully covariant
general two-nucleon mechanism where in the initial collision of the 
projectile with one of the target nucleons, $N^*$(1710), $N^*$(1650),
and $N^*$(1720) baryonic resonances are excited which subsequently 
propagate and decay into the relevant channel. 
Wave functions of baryonic bound states are obtained by 
solving the Dirac equation with appropriate scalar and vector potentials.

In the case of the nucleon projectile,  the cross sections are dominated
by the graphs of the type shown in Fig.1b (target emission graph).
The one-pion-exchange processes make up most of the differential cross
section at all angles. The calculated cross sections are maximum for the
hypernuclear state with the largest orbital angular momentum. For heavier
targets, the angular distributions for the favored transitions peak at angles
larger than the $0^\circ$ which in contrast to the results of most of
the previous nonrelativistic calculations for this reaction. This reflects
directly the nature of the Dirac spinors for the bound states which involve
several maxima in the region of large momentum transfer.
In case of the light target $^4$He, however, the differential cross section
still peaks near the zero degree as in this case the momentum transfers are in
the region where the bound state spinors are smoothly decreasing with
momentum. The energy dependence of the calculated total production cross
section follows closely that of the $pp \to p\Lambda K^+$ reaction.
We find that the nuclear medium corrections to the intermediate
pion propagator introduce large effects on the kaon differential
cross sections. There is also the sensitivity of the cross sections
to the short-range correlation parameter $g^\prime$ in the pion self-energy.
Thus, $(p,K^+)$ reactions may provide an interesting
tool to investigate the medium corrections on the pion propagation
in nuclei.
 
In case of the pion induced reaction, the hypernuclear states with 
stretched spin configurations are preferentially excited.
\begin{figure}
\resizebox{0.35\textwidth}{!}{
\includegraphics{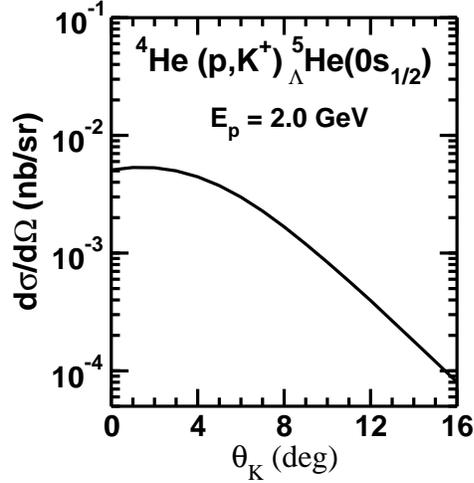}
}
\caption{Differential cross section for the
$^{4}$He$(p,K^+)$$^{5}\!\!\!_\Lambda$He
reaction for the incident proton energy of 2.0 GeV for the
bound state of final hypernucleus as indicated in the figure.
}
\end{figure}

The author wishes to thank Horst Lenske and Ulrich Mosel for several 
useful discussions and collaboration on this subject.

%

\end{document}